\documentclass[aps,preprint,groupedaddress,showpacs,floatfix]{revtex4}
\usepackage{graphicx}
\begin{document}

\title{How learn the branching ratio $X(3872)
\to D^{*0}\bar D^0 + c.c.$ }
\author {
N.N. Achasov$^{\,a}$ \email{achasov@math.nsc.ru} and E.V.
Rogozina$^{\,a,b}$ \email{rogozina@math.nsc.ru}}

\affiliation{
   $^a$Laboratory of Theoretical Physics,
 Sobolev Institute for Mathematics, 630090, Novosibirsk, Russia\\
$^b$Novosibirsk State University, 630090, Novosibirsk, Russia}

\date{\today}

\begin{abstract}
 Enfant terrible of  charmonium spectroscopy, the
resonance $X(3872)$,  generated a stream of interpretations and
ushered in a new exotic $XYZ$ spectroscopy. In the
 meantime, many (if not all)  characteristics of
$X(3872)$ are rather ambiguous. We construct  spectra of decays of
the resonance $X(3872)$  with good analytical and unitary
properties which allows to define the branching ratio of the
$X(3872) \to D^{*0}\bar D^0 + c.c.$ decay studying only one more
decay, for example, the $X(3872)\to\pi^+\pi^- J/\psi(1S)$ decay.
We next define the range of values of the coupling constant of the
$X(3872)$ resonance with the $D^{*0}\bar D^0$ system. Finally, we
show that our spectra are  effective means of selection of models
for the resonance $X(3872)$.
\end{abstract}


\maketitle

\section{Introduction}

 Discovery  of the $X(3872)$ resonance became
the first in discovery of the resonant structures $XYZ$
($X(3872)$, $Y(4260)$, $Z_b^+(10610)$, $Z_b^+(10650)$,
$Z_c^+(3900)$), the resonant interpretations of which assumes
existence in them at least pair of heavy and pair of light quarks
in this or that form. Thousand articles on this subject already
were published in spite of the fact that many properties of new
resonant structures are not defined yet and not all possible
mechanisms of dynamic generation of these structures are studied,
in particular, the role of  the anomalous Landau thresholds is not
studied.

Below we suggest an approach which allows to define the branching
ratio of the $X(3872) \to D^{*0}\bar D^0 + c.c.$ decay studying
only one more decay of $X(3872)$ into a non-$D^{*0}\bar D^0$
channel and to select models the $X(3872)$ resonance.

The paper is organized as follows. In Sec. 2 we construct spectra
of decays of the resonance $X(3872)$  with good analytical and
unitary properties and define the range of values of the coupling
constant of the $X(3872)$ resonance with the $D^{*0}\bar D^0$
system.  In Sec. 3 we show that the constructed spectra can
effectively select the model proposed for the $X(3872)$ resonance.

\section{Spectra and coupling constant}
\subsection{The $X(3872) \to D^{*0}\bar D^0 + c.c.$ and others
spectra}

\begin{figure}[h]
\begin{center}
\begin{tabular}{ccc}
\includegraphics[width=8.5cm,height=5.9cm]{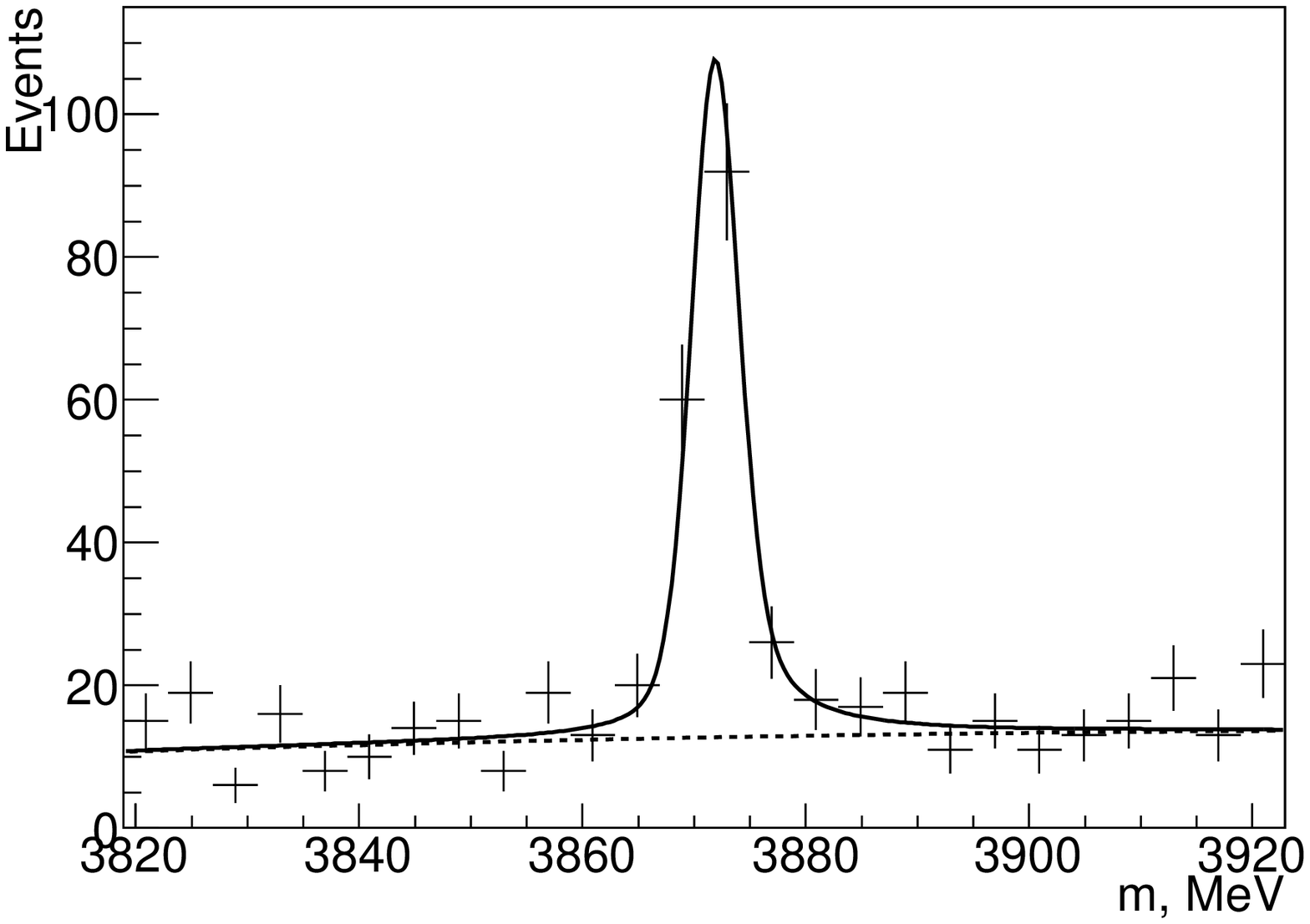}& \includegraphics[width=8.5cm,height=5.9cm]{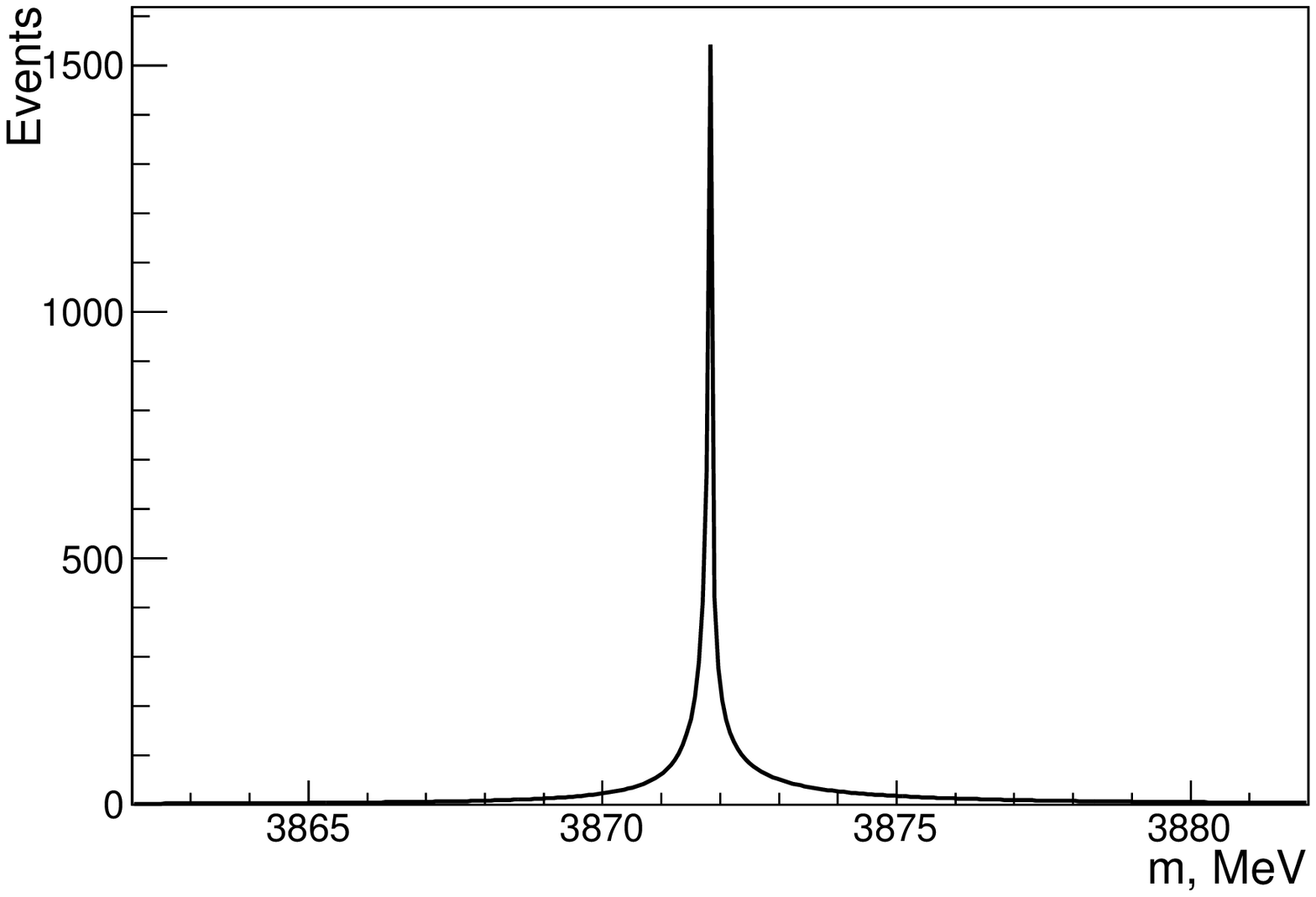}\\ (a)&(b)
\end{tabular}
\end{center}
\caption{a) The Belle data \cite{Belle11} on the invariant
$\pi^+\pi^- J/\psi(1S)$ mass ($m$) distribution. The solid line is
our theoretical one with taking into account the Belle energy
resolution. The dotted line is second-order polynomial for the
incoherent background. b) Our undressed theoretical line. }
\label{fig1}
\end{figure}

 The mass spectrum
$\pi^+\pi^-J/\psi(1S)$ in the $X(3872)\to\pi^+\pi^-J/\psi(1S) $
decay \cite{Belle11} looks as the ideal Breit-Wigner one, see Fig.
\ref{fig1}a.

 The mass spectrum $\pi^+\pi^-\pi^0 J/\psi(1S)$ in
the $X(3872)\to\pi^+\pi^-\pi^0 J/\psi(1S)$ decay
 looks in a similar way \cite{Belle05,BABAR10}.

\begin{figure}[h]
\begin{center}
\begin{tabular}{ccc}
\includegraphics[width=8.5cm,height=6cm]{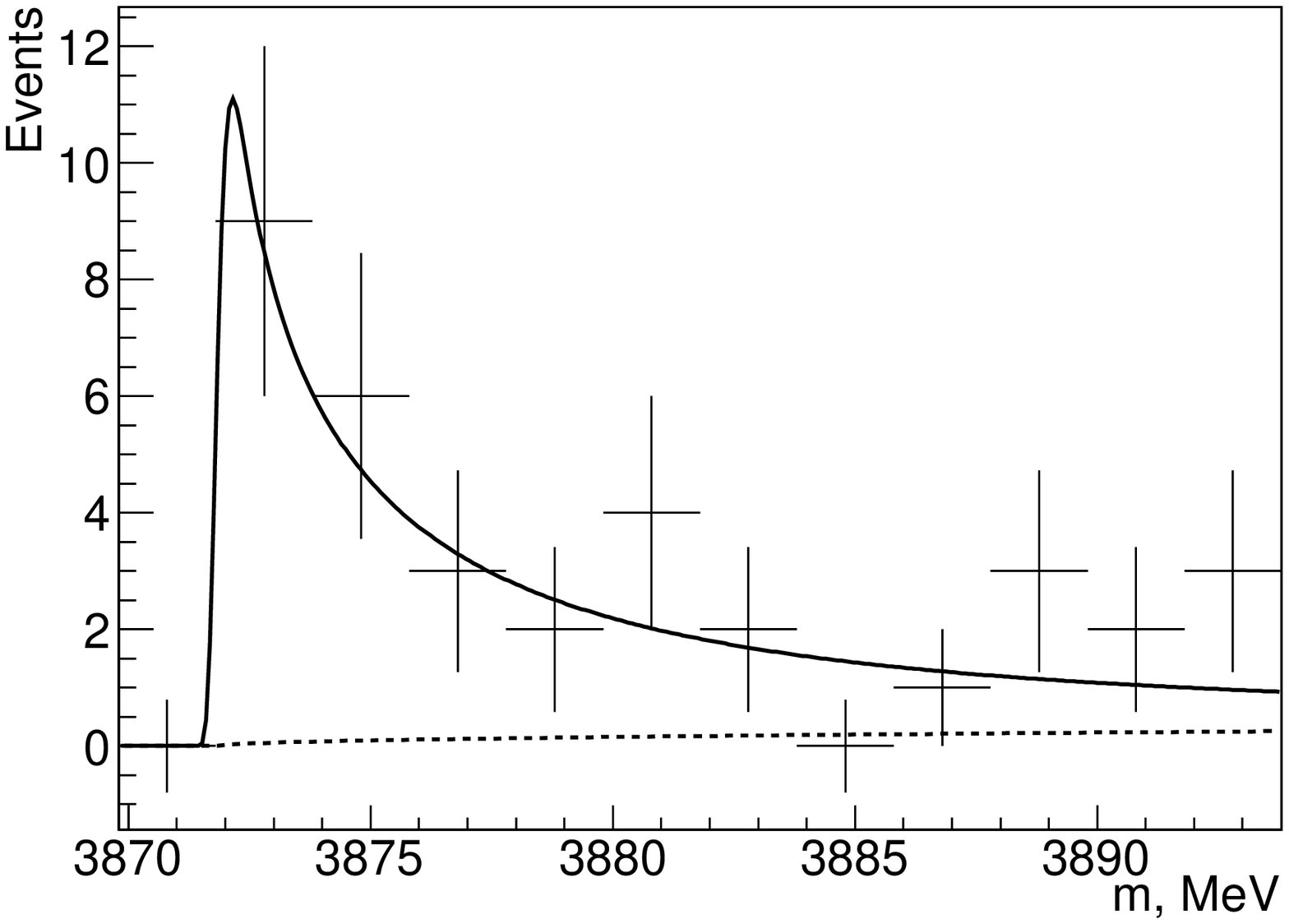}& \includegraphics[width=8.5cm,height=6cm]{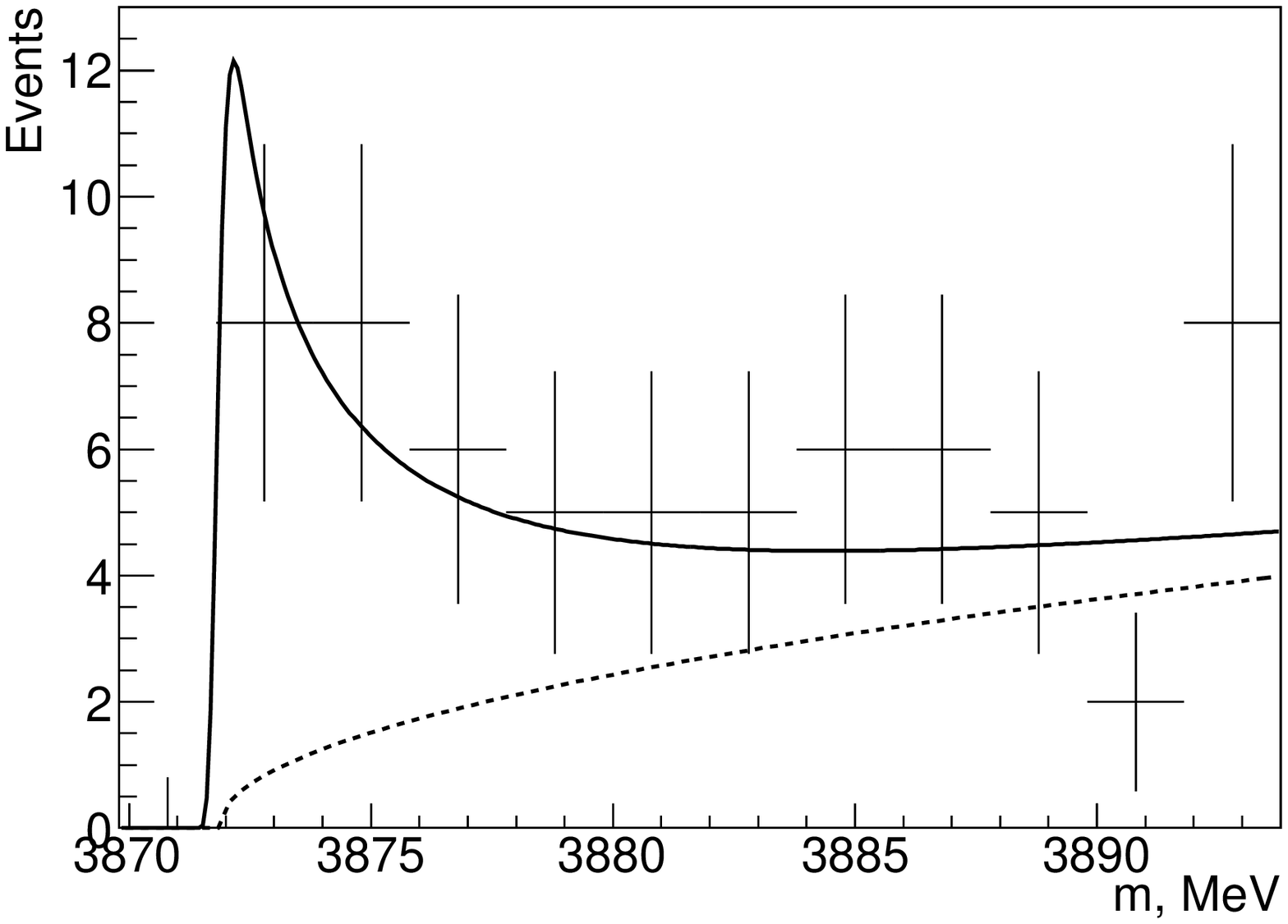}\\ (a)&(b)
\end{tabular}
\end{center}
 \vspace*{-15pt}
\caption{The Belle data \cite{Belle10} on the invariant
$D^{*0}\bar D^0 + c.c.$ mass ($m$) distribution. The solid line is
our theoretical one with taking into account the Belle energy
resolution. The dotted line is a square root function for the
incoherent background. a)  $D^{*0}\to D^0\pi^0$. b) $D^{*0}\to
D^0\gamma$. } \label{fig2}
\end{figure}

The mass spectrum $D^{*0}\bar D^0 + c.c.$ in the $X(3872)\to
D^{*0}\bar D^0 + c.c.$ decay \cite{Belle10} looks as the typical
resonance threshold enhancement, see Fig. \ref{fig2}. \footnote{An
interference between the $D^0\bar D^{*0}$ and $\bar D^0 D^{*0}$
channels is negligible for the narrowness of the $D^{*0}$ and
$\bar D^{*0}$ states. Using the isotopical invariance of the
$DD^*\pi$ interacion and the experimental information about the
$D^{*+}\to D^0\pi^+$ and $D^{*+}\to D^+\pi^0$ decays, one can find
$\Gamma_{D^{*0}}= 70$ KeV.}

 If structures in the above channels are manifestation of the same resonance, it is possible to  define the branching ratio
$X(3872) \to D^{*0}\bar D^0 + c.c.$, $BR(X(3872) \to D^{*0}\bar
D^0 + c.c.)$ treating data only these (two) decay channels.

We believe that the $X(3872)$ is the axial vector, $1^{++}$
\cite{LHCb,PDG12}. In this case the S wave dominates in the
$X(3872) \to D^{*0}\bar D^0 + c.c.$ decay and hence is described
by the effective Lagrangian
\begin{equation}
\label{L} L_{XD^{*0}D^0}(x)=g_AX^\mu\Bigl(D^0_\mu(x)\bar D^0(x)+
\bar D^0_\mu(x)D^0(x)\Bigr ).
\end{equation}
The width of the $X\to D^{*0}\bar D^0 + c.c.$ decay
\begin{equation}
\label{width}
 \Gamma(X \to D^{*0}\bar D^0 + c.c.\,,\, m
 )=\frac{g_A^2}{8\pi}\frac{\rho(m)}{m}\Biggl(1+\frac{{\bf k}^2}{3m_{D^{*0}}^2}\Biggr)\,,
 \end{equation}
where $\bf k$ is momenta of $D^{*0}$ (or $\bar D^0$) in the
$D^{*0}\,\bar D^0$ center mass system, $m$ is the invariant mass
of the $D^{*0}\,\bar D^0$ pair,
\begin{equation}
\label{rho} \rho(m)=\frac{2|{\bf k}|}{m}=
\frac{\sqrt{(m^2-m_+^2)(m^2-m_-^2)}}{m^2}\,,\ \ \ \ m_\pm
=m_{D^{*0}}\pm m_{D^0}\,.
\end{equation}

The second term in the right side of Eq. (\ref{width}) is very
small in our energy region and can be neglected. This gives us the
opportunity to construct the mass spectra for the  $X(3872)$
decays with the good analytical and unitary properties as in the
scalar meson case \cite{ads,nna-avk}.

The mass spectrum in the $D^{*0}\bar D^0 + c.c.$ channel
\begin{equation}
\label{SpectrumDD*} \frac{ dBR(X\to D^{*0}\bar D^0+c.c.\,,\, m
)}{dm} =4\frac{1}{\pi}\frac{m^2\Gamma(X\to D^{*0}\bar D^0,\,
m)}{|D_X(m)|^2}.
\end{equation}

The branching ratio of $X(3872) \to D^{*0}\bar D^0 + c.c.$
\begin{equation}
\label{BRDD*}
 BR(X\to
D^{*0}\bar D^0+c.c.) = 4\frac{1}{\pi}\int_{m_+}^\infty
\frac{m^2\Gamma(X\to D^{*0}\bar D^0,\, m)}{|D_X(m)|^2} dm\,.
\end{equation}

In others $\{i\}$ (non-$D^{*0}\bar D^0$) channels the $X(3872)$
state is seen as a narrow resonance that is why we write the mass
spectrum in the $i$ channel in the form
\begin{equation}
\label{Spectrumi}
 \frac{dBR(X\to i\,,\, m )}{dm}
=2\frac{1}{\pi}\frac{m_X^2\,\Gamma_i}{|D_X(m)|^2}\,,
\end{equation}
where $\Gamma_i$ is the width of the $X(3872)\to i$ decay.

 The branching ratio of $X(3872)\to i$
\begin{equation}
\label{BRi}
 BR(X\to i) = 2\frac{1}{\pi}\int_{m_0}^\infty
\frac{m_X^2\Gamma_i}{|D_X(m)|^2} dm\,,
\end{equation}
where $m_0$ is the threshold of the $i$ state.
\begin{equation}
\label{inverseprop}
 D_X(m)= m_X^2-m^2 + Re(\Pi^{D^{*0}\bar D^0}_X(m_X))-
\Pi^{D^{*0}\bar D^0}_X(m)-\imath
 m_X\Gamma\,,
 \end{equation}
where $\Gamma=\Sigma\Gamma_i$ is the total width of the $X(3872)$
decay into all non-$D^{*0}\bar D^0$ channels.

When  $m_+ \leq m$,
\begin{equation}
\label{mm+}
 \hspace*{-21pt} \Pi^{D^{*0}\bar D^0}_X(m)=\frac{g_A^2}{8\pi^2}\left\{
\frac{(m^2-m_+^2)}{m^2}\frac{m_-}{m_+}\ln\frac{m_{D^{*0}}}{m_{D^0}}
 +\rho(m)\left [\imath\pi +
\ln\frac{\sqrt{m^2-m_-^2}-\sqrt{m^2-m_+^2}}{\sqrt{m^2-m_-^2}+\sqrt{m^2-m_+^2}}
\right ]\right\}.
\end{equation}

When  $m_- \leq m\leq m_+$,
\begin{equation}
\label{m-mm+} \Pi^{D^{*0}\bar
D^0}_X(m)=\frac{g_A^2}{8\pi^2}\left\{
\frac{(m^2-m_+^2)}{m^2}\frac{m_-}{m_+}\ln\frac{m_{D^{*0}}}{m_{D^0}}
 -2|\rho(m)|\arctan\frac{\sqrt{m^2-m_-^2}}{\sqrt{m_+^2-m^2}}\right\},
\end{equation}
where $|\rho(m)|= \sqrt{(m_+^2-m^2)(m^2-m_-^2)}/m^2$.

When $m\leq m_-$ and $m^2\leq 0$,
\begin{equation}
\label{mm-}
 \Pi^{D^{*0}\bar D^0}_X(m)=\frac{g_A^2}{8\pi^2}\left\{
\frac{(m^2-m_+^2)}{m^2}\frac{m_-}{m_+}\ln\frac{m_{D^{*0}}}{m_{D^0}}
 -\rho(m)\ln\frac{\sqrt{m_+^2-m^2}-\sqrt{m_-^2-m^2}}{\sqrt{m_+^2-m^2}+\sqrt{m_-^2-m^2}}
\right\}.
\end{equation}

Our branching ratios satisfy unitarity
 \begin{equation}
 \label{unitarity}
 1= BR(X\to D^{*0}\bar D^0+c.c.)+\sum_iBR(X\to i)\,.
 \end{equation}

Fitting the Belle data \cite{Belle11,Belle10}, we take into
account the Belle \cite{Belle11} results that $m_X=
3871.84\,\mbox{MeV}= m_{D^{*0}}+ m_{D^0}= m_+$ and
$\Gamma_{X(3872)}<1.2$ MeV 90\%CL that corresponds to $\Gamma
<1.2$ MeV,
 which controls
the width of the $X(2872)$ signal in the $\pi^+\pi^- J/\psi(1S)$
channel and in every non-$D^{*0}\bar D^0$
 channel, see Fig. \ref{fig1}b.

 The results of our fit are  in the
 Table I. The current statistics is not sufficient for  serious
 conclusions. \\ [12pt]
  TABLE I. Results of the analysis of the Belle
data \cite{Belle11,Belle10}.\\ $BR_{seen}=BR(X\to D^{*0}\bar
D^0+c.c.\,; m\leq 3891.84\,\mbox{MeV})$, $BR=BR(X\to D^{*0}\bar
D^0+c.c.)$,\\ $BR(Oth)_{seen}=\sum_iBR(X\to i\,;\, 3851.84\leq
m\leq 3891.84\,\mbox{MeV})$. $\Gamma$ in MeV, $g_A$ in GeV.\\[9pt]
\renewcommand{\arraystretch}{1.2}
\renewcommand{\tabcolsep}{3 mm}
\begin{tabular}{cccccc}
\hline \hline $\Gamma$ & $g_A^2/{8\pi}$ & $\chi^2/Ndf$ &
$BR_{seen}$ &  $BR$ & $BR(Oth)_{seen}$\\ \hline $1.2_{-0.467}$ &
$0.857_{-0.481}^{+3.614}$ & 43.74/42 & $0.486_{-0.29}^{+0.061}$ &
$0.795_{-0.224}^{+0.19}$ & $0.191_{-0.179}^{+0.223}$\\ \hline
\hline
\end{tabular}
\\[24pt]
Nevertheless, one can state that our results are consist with
experiment.  Really, in view of \\ $BR(B\to X(3872)K)\times
BR(X(3872)\to D^{*0}\bar D^0) = (0.80\pm 0.20\pm 0.1)\times
10^{-4}$ \cite{Belle10},\\ $BR(B^+\to X(3872)K^+)\times
BR(X(3872)\to\pi^+\pi^- J/\psi(1S)) = (8.61\pm 0.82\pm 0.52)\times
10^{-6}$ \cite{Belle11},\\ $BR(B^+\to X(3872)K^+)\times
BR(X(3872)\to\pi^+\pi^-\pi^0 J/\psi(1S)) = (0.6\pm 0.2\pm
0.1)\times 10^{-5}$ \cite{BABAR10},\\ and $BR(B^+\to
X(3872)K^+)\times BR(X(3872)\to\gamma J/\psi(1S)) =
(1.78^{+0.48}_{-0.44}\pm 0.12)\times 10^{-6}$ \cite{Belle11g}\\ it
follows that $BR(X\to D^{*0}\bar D^0+c.c.\,; m\leq
3892\,\mbox{MeV})$  is a few times as large as the sum of all
non-$D^{*0}\bar D^0$ known branching ratios.

 So, when  fitting the $X(3872)\to D^{*0}\bar D^0$
data and data for any $X(3872)$ decay  into  non-$D^{*0}\bar D^0$
state, $X(3872)\to i $, we find $\Gamma$ and $g_A^2/8\pi$, which
define $BR(X(3872)\to D^{*0}\bar D^0+c.c.)$. Generally speaking,
we don't need to know $BR(X(3872)\to i)$.

\subsection{Influence of the $X(3872) \to D^{*+}\bar D^- + c.c.$
channel}

As seen from Table I the sizeable part (near 40\%) of $BR=BR(X\to
D^{*0}\bar D^0+c.c.)$ accounts for the tail of the $X(3872)$
resonance ($m\geq 3891.84$ MeV). This gives an idea to take into
account the  $X(3872)\to D^{*+}D^- +c.c. $ decays \footnote{An
interference between the $D^+ D^{*-}$ and $D^- D^{*+}$ channels is
negligible for the narrowness of the $D^{*\pm}$ states,
$\Gamma_{D^{*\pm}}= 96$ KeV.} on the $X(3872)$ tail. Since
$X(3872)$ is an isoscalar, the effective Lagrangian has the form
\begin{equation}
\label{L2} L(x)=g_AX^\mu\Bigl(D^0_\mu(x)\bar D^0(x)+ \bar
D^0_\mu(x)D^0(x)+ D^+_\mu(x)D^-(x)+ D^+_\mu(x)D^-(x)\Bigr ).
\end{equation}

Eq. \ref{inverseprop} is replaced by
 \begin{equation}
\label{inverseprop2}
 D_X(m)= m_X^2-m^2 + Re(\Pi_X(m_X))-
\Pi_X(m)-\imath  m_X\Gamma\,,
 \end{equation}
where
\begin{equation}
\label{Pi(m)}
 \Pi_X(m)= \Pi^{D^{*0}\bar D^0}_X(m)+ \Pi^{D^{*+}D^-}_X(m)\,.
 \end{equation}
$\Pi^{D^{*+}D^-}_X(m)$ is obtained from $\Pi^{D^{*0}\bar
D^0}_X(m)$, see Eqs. (\ref{mm+}), (\ref{m-mm+}) and (\ref{mm-}),
by replacement of $m_{ D^{*0}}$ and $m_{ D^0}$ by $m_{ D^{*+}}$
and $m_{ D^+}$, respectively.

The unitarity condition, Eq. (\ref{unitarity}), takes the form
 \begin{equation}
 \label{unitarity2}
 1= BR(X\to D^{*0}\bar D^0+c.c.)+ BR(X\to D^{*+}\bar D^-+c.c.) + \sum_iBR(X\to i)\,.
 \end{equation}

 The results of our fit are  in the
 Table II.\\[15pt]
TABLE II.  $\Gamma$ in MeV, $g_A$ in GeV.
\\[9pt]
\begin{tabular}{|cc|cccc|}
\hline $\Gamma$ & $1.2_{-0.42}$ & mode & $X\to D^{*0}\bar
D^0+c.c.$ & $X\to D^{*+}D^{-}+c.c.$ & $X\to Others$\\ \hline
$g_A^2/{8\pi}$ & $1.36_{-0.95}^{+4.85}$ & $BR$ &
$0.586_{-0.101}^{+0.025}$ & $0.315_{-0.16}^{+0.132}$ &
$0.098_{-0.096}^{+0.261}$\\ \hline $\chi^2/Ndf$ & 45.49/42 &
$BR_{seen}$ & $0.285_{-0.188}^{+0.121}$ &
$0.028_{-0.019}^{+0.004}$ & $0.091_{-0.084}^{+0.255}$\\ \hline
\end{tabular}
\\[30pt]
The results in Tables I and II are compatible within the errors.
The corresponding curves are similar to ones  in Figs. \ref{fig1}
and \ref{fig2}. Of course, one should take into account the
$X(3872)\to D^{*+}D^- +c.c. $ channel in the case of the good
statistics.

\section{conclusion}
 Our approach can serve as the guide in
selection of theoretical models for the $X(3872)$ resonance.
Indeed, if $3871.68$ MeV $<m_X<3871.95$ MeV \cite{PDG12} and
$\Gamma_{X(3872)}=\Gamma <1.2$ MeV \cite{PDG12} then  for
$g_A^2/8\pi<0.2$ GeV$^2$ (that does not contradict current
experiment, see Tables I and II) $BR(X\to D^{*0}\bar
D^0+c.c.)=BR<0.3$. That is,
 unknown
decays of $X(3872)$ into non-$D^{*0}\bar D^0$ states are
considerable or dominant.

For example, in Ref. \cite{Maiani} the authors considered
$m_X=3871.68$ MeV, $\Gamma=1.2$ MeV and
$g_{XDD^*}=g_A\sqrt{2}=2.5$ GeV, that is, $g_A^2/8\pi=0.1$
GeV$^2$. In this case $BR(X\to D^{*0}\bar D^0+c.c.)\approx 0.2$,
that is, unknown decays $X(3872)$ into non-$D^{*0}\bar D^0$ states
are dominant. For details see Table III.\\[33pt]
 TABLE III. Branching ratios for the model from Ref. \cite{Maiani} without $D^{*+}D^{-}+c.c.$ channel.
 $m_X$ in MeV, $\Gamma$ in MeV, $g_A$ in GeV. \\[9pt]
\begin{tabular}{cccccc}
\hline \hline $m_X$ & $\Gamma$ & $g_A^2/{8\pi}$ & $BR_{seen}$ &
$BR$ & $BR(Oth)_{seen}$\\ \hline $3871.68$ & $1.2$ & $0.1$ &
$0.152$ & $0.189$ & $0.792$\\ \hline \hline
\end{tabular}
\\[30pt]
Account of influence of the $X(3872) \to D^{*+}\bar D^- + c.c.$
channel has little effect on the results for the model from Ref.
\cite{Maiani}  since $g_A^2/{8\pi}$ is small , see Table
IV.\\[15pt]
 TABLE IV. Branching ratios for the model from Ref.
\cite{Maiani}. $m_X$ in MeV, $\Gamma$ in MeV, $g_A$ in GeV.\\[9pt]
\begin{tabular}{|cc|cccc|}
\hline
 $m
 _X$ & $3871.68$ & mode & $X\to D^{*0}\bar D^0+c.c.$ & $X\to D^{*+}D^{-}+c.c.$ & $X\to Others$\\
\hline $\Gamma$ & $1.2$ & $BR$ & $0.176$ & $0.045$ & $0.779$\\
\hline $g_A^2/{8\pi}$ & $0.1$ & $BR_{seen}$ & $0.14$ & $0.011$ &
$0.761$\\ \hline
\end{tabular}

\vspace*{30pt}

{\bf Acknowledgments.}

 We are grateful to T.A.-Kh. Aushev and G.V. Pakhlova for the
consultation. This work was supported in part by RFBR, Grant No
13-02-00039, and Interdisciplinary project No 102 of Siberian
division of RAS.\\[9pt]

\end{document}